\documentclass{article}
\usepackage{graphicx} 
\usepackage{amssymb}
\usepackage{pifont}
\usepackage{enumitem}
\usepackage{tikz}
\usepackage{xcolor}
\usepackage{geometry}

\usepackage{amsmath} 
\usepackage{hyperref}
\usepackage{float}
\usepackage{tabularx} 
\usepackage{gensymb}
\usepackage{textcomp}
\usepackage{caption}
\usepackage{subcaption}
\usepackage{array}
\usepackage{multirow}
\usepackage{listings}
\usepackage[dvipsnames]{xcolor}
\usepackage{pythonhighlight}
\usepackage{tocloft}
\usepackage{fancyhdr}
\usepackage{mathtools}
\usepackage{pgfplots}
\usepackage{changepage}
\usepackage{pgfplotstable}
\usepackage{tipa}
\usepackage{setspace}
\usepackage{soul}
\usepackage{capt-of}
\usepackage{afterpage}
\usepackage{xcolor}
\usepackage[percent]{overpic}
\usepackage{authblk}
\usepackage{etoolbox}
\makeatletter
\patchcmd{\@bibitem}
  {\newblock}
  {\newblock\bfseries}
  {}{}
\makeatother

\usepackage[super]{natbib} 
\bibpunct{}{}{,}{s}{}{}

\graphicspath{{figures/}}

\pgfplotsset{compat=1.18}
\fontsize{12pt}{12pt}\selectfont
\begin{document}
\title{Shock-induced tipping in a thermoacoustic system}

\author[1,2]{Bhadra Sreelatha}
\author[1,2]{Rohit Radhakrishnan}
\author[1,2]{R. I. Sujith\thanks{sujith@iitm.ac.in}}

\affil[1]{Department of Aerospace Engineering, Indian Institute of Technology Madras}
\affil[2]{Center of Excellence for studying Critical Transitions in Complex Systems}
\date{}
\maketitle

\section*{Abstract}

\large Tipping refers to the transition of a system from one state to another. In this study, we focus on shock-induced tipping, which occurs due to a sudden and large disturbance in a control parameter, which is referred to as the shock. This shock drives the system from one dynamical state to another. We present the first experimental demonstration of shock-induced tipping using a prototypical thermoacoustic system, the horizontal Rijke tube. In a thermoacoustic system, unsteady heat release and sound waves interact through positive feedback, leading to self-sustained, high amplitude oscillations known as limit cycles. The system transitions from a quiescent state to a state of self-sustained oscillations when a shock is introduced in the power supplied to the heat source (an electrically heated grid). This shock is created by abruptly increasing the voltage supplied to the grid, which takes the system into a bistable region. To explain the underlying mechanism linking the shock in the supplied power to the observed tipping behavior, we model the system by modifying the governing equations of the Rijke tube to incorporate the heat transfer properties of the grid. We demonstrate that the shock in the supplied power manifests as a shock in the grid temperature, causing the system to fall into the basin of attraction of an alternate stable state. The tipping event depends on the magnitude of shock and temperature of the grid. Understanding the mechanisms underlying shock-induced tipping is crucial for developing systems with improved safety and reliability.\\

\section{Introduction} \label{intro}
\large
Tipping describes a critical transition where a system undergoes a sudden and often irreversible shift to a different state~\cite{vanNes2016tipping}. Such transitions are observed in biological systems~\cite{venegas2005asthma}, ecological systems~\cite{scheffer2001catastrophic}, economical systems~\cite{johansen1997crashes} and climate systems~\cite{tmlentontipping}. Ashwin et al.~\cite{ashwin2012tipping} outlined three mechanisms for tipping: namely bifurcation-induced (B-tipping), noise-induced tipping (N-tipping) and rate-induced tipping (R-tipping).
\\
 \large B-tipping occurs when a gradual change in a system parameter pushes the system past a critical point, causing a shift to a different state~\cite{ashwin2012tipping,EAGopalakrishnanInfluence}. R-tipping occurs when the system is unable to track its quasi-static attractor because the rate of variation of a control parameter exceeds a critical value, leading to a transition to an alternate state.~\cite{ashwin2012tipping,pavithran2023tipping,pavithran2021rate,JTonypreconditionedrate,feudel2023rate}. N-tipping occurs due to stochastic fluctuations in the system that destabilize a state, pushing it toward a different equilibrium~\cite{ashwin2012tipping,unni2019interplay}.
 These three modes of tipping have been extensively observed and analyzed in a variety of systems. A fourth theoretically proposed tipping mechanism, known as shock-induced tipping~\cite{feudel2023rate,halekotte2020minimal,bifurcationdiagramsincomplexclimate} (S-tipping), occurs when a single large disturbance causes the system to tip to fall into the basin of attraction of another state. S-tipping can be considered as the limiting case of R-tipping in which the rate of change of parameter approaches infinity. While rate-induced tipping arises from finite-time parameter variations whose impact depends on both their magnitude and duration in which it is applied, shock tipping occurs when the forcing is applied abruptly. In this instantaneous limit~\cite{abottcriticalrate}, the notion of rate loses relevance, and the system tips only if the imposed shock exceeds a critical magnitude~\cite{schefferpulsedrive}. The shock displaces the system out of the basin of attraction of its desired state and into an alternate dynamical state~\cite{halekotte2020minimal}.
\\
\large  In R-tipping, the control parameter changes continuously from the beginning at a specified  rate. The tipping occurs because the system cannot adapt sufficiently quickly to the steadily changing parameter, even though no abrupt external shock is applied~\cite{pavithran2023tipping,pavithran2021rate}. In contrast, S-tipping is a mechanism that involves a sudden and large disturbance that manifests itself as a sudden increase in the rate of the control parameter~\cite{feudel2023rate}. Although the parameter may have been varying at a uniform rate beforehand, an abrupt disturbance can push the system beyond the boundary of its current basin of attraction, triggering a transition to a different state.~\cite{feudel2023rate,halekotte2020minimal}. Although experimental evidence is yet to be reported, this type of tipping can be particularly concerning, as it can occur unexpectedly without any warning and lead to catastrophic events. 
\\
 \large The occurrence of S-tipping is conceptually identified in some real systems. In the Great Britain power grid~\cite{halekotte2020minimal} for example, S-tipping can occur due to sudden disturbances such as a generator failure or a sudden surge in demand. These disturbances can overload critical transmission lines, particularly in weakly connected regions, causing parts of the grid to fall out of sync and potentially leading to widespread desynchronization or blackouts. Similar dynamics are observed in ecological systems. For instance, plant pollinator networks are theoretically  shown to undergo S-tipping if a key species is abruptly removed, triggering cascading failures that ultimately lead to the collapse of the entire network~\cite{halekotte2020minimal}. 
However, despite its relevance in real-world contexts, studies in S-tipping has not been experimentally demonstrated and remains largely theoretical.
\\
\large In the present study, we investigate the phenomenon of S-tipping using a prototypical thermoacoustic system, a horizontal Rijke tube. Thermoacoustic systems involve the interaction between acoustic waves and heat source which result in a positive feedback and lead to the onset of self sustained limit cycle oscillations~\cite{balasubramanian2008thermoacoustic,subramanian2010bifurcation}. The Rijke tube provides various advantages such as the capacity to perform highly controlled experiments, supports high data sampling frequency and allows the application of different types of control parameter changes~\cite{unni2019interplay,etikyala2017change,JTonypreconditionedrate}.
\\
 \large As the control parameter is varied, the Rijke tube can exhibit transitions from a non-oscillatory state to a state of self-sustained, high amplitude periodic oscillations. This oscillatory behaviour is commonly referred to in the literature~\cite{juniper2018sensitivity,RISujithTextbook,Smariappan-modellingnonlineartai} as thermoacoustic instability (TAI). The state of TAI is referred in many systems such as propulsion systems of rockets~\cite{fisher2009apollo} and gas turbines~\cite{lieuwen2005combustion}. Transitions into such states can be caused by sudden disturbances in the systems through S-tipping. Sudden unexpected changes in system parameters can arise inadvertently; however these changes have the potential to take the system into dangerous states without the control parameter even crossing the stability boundaries of the system.  
This paper aims to provide experimental evidence of S-tipping and provides an understanding of the mechanism behind such tipping in thermoacoustic systems. We develop a mathematical model to understand how the sudden and large change in the control parameter is causing the system to tip into LCO through S-tipping.
\\
\large The structure of the paper is as follows: Section~\ref{exp_setup} provides a detailed account of the experimental setup used in the study.  Next, in Section~\ref{exp_results}, we present the experimental observations and key findings. Section~\ref{math_model} introduces the mathematical model developed to describe the behavior of the system. Section~\ref{modelresults} presents the results of the mathematical model which illustrates the proposed mechanism of the transition of the system from quiescent operation into oscillatory states through S-tipping in our thermoacoustic system.
\graphicspath{{figures/}}
\section{Experimental setup} \label{exp_setup}
\begin{figure}[H]
    \centering
    
    \begin{subfigure}[b]{0.51\textwidth}
        \centering
        \includegraphics[width=\textwidth]{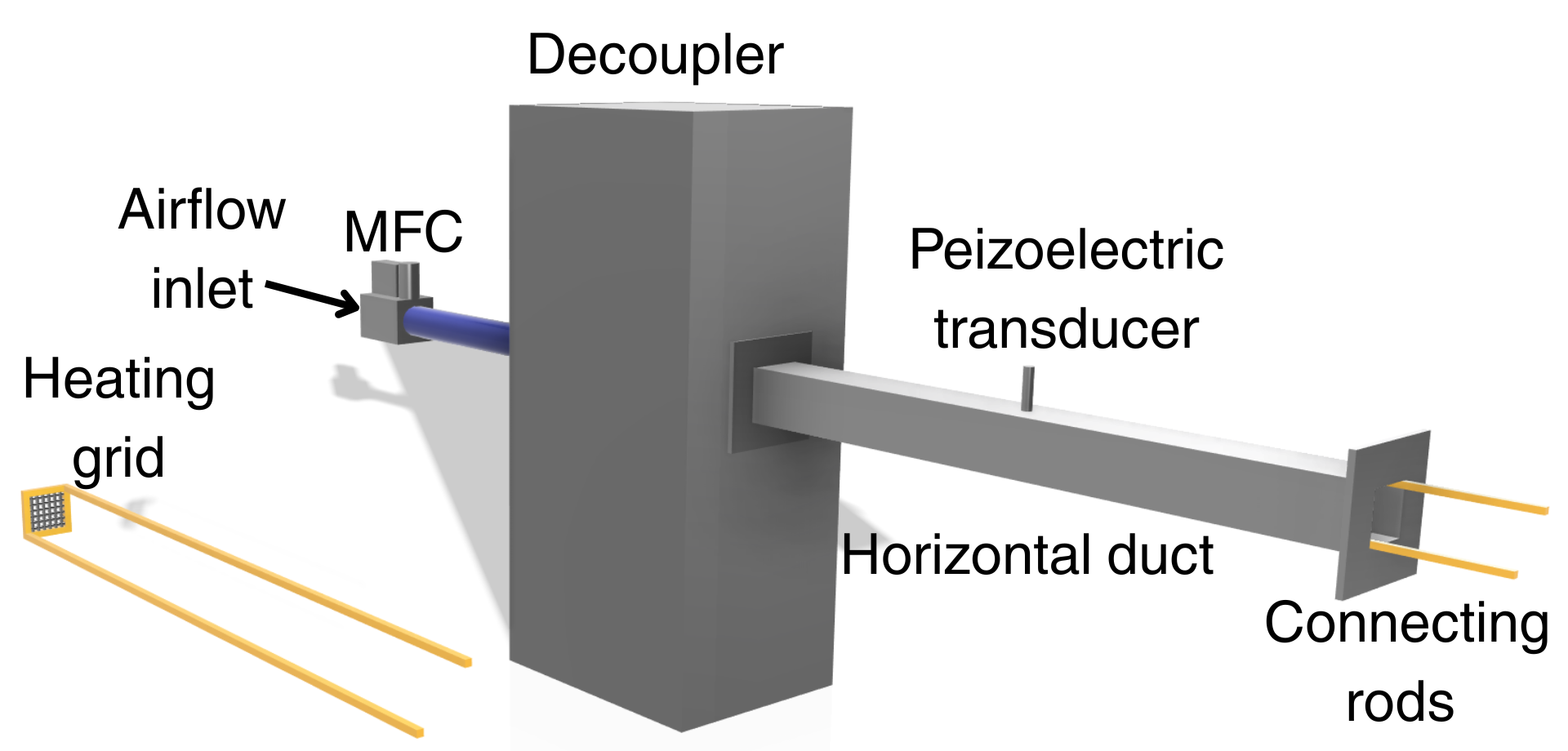}
        \caption{}
        \label{setup}
    \end{subfigure}
    \hfill
    \begin{subfigure}[b]{0.48\textwidth}
        \centering
        \includegraphics[width=\textwidth]{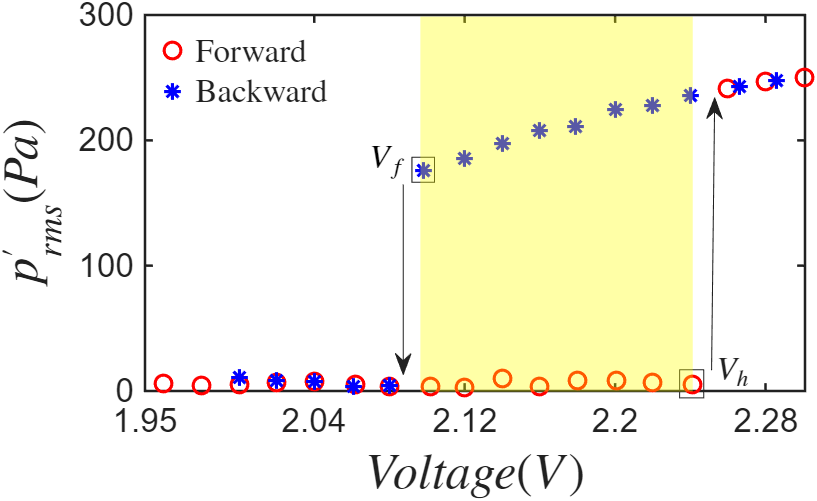}
        \caption{}
        \label{expbifplot}
    \end{subfigure}

   \caption{\large Schematic of the horizontal Rijke tube, comprising of a long square duct, decoupler, mass flow controller, connecting rods supplying current to an electrically heated grid, and a piezoelectric pressure transducer for measuring acoustic signals. (b) Bifurcation diagram constructed by plotting the RMS value of acoustic pressure oscillations against the the voltage supplied to the grid. The bistable region of the system is represented in yellow.}
\end{figure}

\large 
  We perform this study on a laminar thermoacoustic system, the horizontal Rijke tube~\cite{balasubramanian2008thermoacoustic,subramanian2010bifurcation,matveev2003thesis,EAGopalakrishnanInfluence}. Figure~\ref{setup} illustrates the configuration of the experimental setup. The Rijke tube consists of a horizontal duct, 1 m in length with a square cross-section of 92 mm  $\times$ 92 mm.  A compressor is used to establish the airflow through the duct, with a mass flow controller (MFC) regulating the flow rate at 100 $\pm$ 1.8 SLPM and a mean bulk velocity, $u_0 = (19.6 \pm 0.3)\times10^{-2} $ m/s. Both ends of the duct are open, with one end exposed to atmospheric conditions and the other opening into a rectangular chamber (1200 mm $\times$ 450 mm $\times$ 450 mm), referred to as the decoupler, which removes any inherent fluctuations present in the inlet flow~\cite{gopalakrishnanews,etikyala2017change}.  The Reynolds number is evaluated for the flow, defined as  $Re = \rho u L/\mu,$ where \( L \) is the characteristic dimension, taken as the width of the duct cross-section (92 mm), \( \rho = 1.2 \) $\text{kg/m}^3$ is the density of air at 297 K and \( \mu = 1.6 \times 10^{-5}\) $\text{kg/m$\cdot$s} $ is the dynamic viscosity of air at 297 K. Substituting these values yields \( Re = 1353 \pm 0.1\), which lies within the laminar regime for flow in a rectangular duct~\cite{song2020flow}.
\\
\large An electrically heated grid made of mild steel, positioned 25 cm from the inlet serves as the heat source for the system. The grid is powered by a programmable DC power supply.  
The electric power supplied to the grid is regulated by controlling the voltage supplied to the grid. A flush-mounted piezoelectric pressure transducer (PCB 103B02, sensitivity of 217.5 mV/kPa and an uncertainty of $\pm 0.2$\,Pa), positioned 50 cm from the upstream end of the duct, measures acoustic pressure fluctuations in the duct. The acquired acoustic pressure fluctuations are captured using a National Instruments USB--6343 data acquisition system.

\large
The exponential decay rate due to the acoustic damping in the system under cold flow is kept within bounds~\cite{etikyala2017change}, to ensure the repeatability of the experiments. 
The decay rate which is maintained at $11.9 \pm 1$ s$^{-1}$ is determined by applying the Hilbert transform to the acoustic pressure signals inside the duct and calculating the logarithmic decay of its amplitude~\cite{etikyala2017change}.
\\
\large The mass flow controller (MFC) is set to establish a flow rate of 100 SLPM for which subcritical Hopf bifurcation is expected in this Rijke tube~\cite{etikyala2017change,EAGopalakrishnanInfluence}. In our system, Hopf bifurcation describes the transition from a quiescent state to self-sustained limit cycle oscillations (LCO) as the power supplied to the grid is increased in a quasi-static manner by controlling the voltage supplied to the grid~\cite{subramanian2010bifurcation}. The transition to LCO occurs at a critical value of the power supplied to the grid, referred to as the Hopf point. As shown in Fig.~\ref{expbifplot}, when the voltage is quasi-statically decreased, the system transitions from the state of LCO, back to a quiescent state at a value of power, known as the fold point.
\\
 
\large At each value of the voltage, the system is allowed to remain undisturbed for 240 s allowing it to attain a stationary state. Following the identification of the voltage corresponding to Hopf and fold points (denoted by $V_h$ and $V_f$ respectively), experiments were conducted wherein the voltage supplied to the grid was varied linearly at a constant rate. In laminar thermoacoustic systems, varying the control parameter at a constant rate can trigger transitions from quiescent operation to oscillatory states at parameter values that differ from the Hopf point due to rate-dependent tipping delay~\cite{pavithran2021rate}.
\\
\large Another kind of control parameter variation which is seen in many systems is when an abrupt increase or a shock is applied to the parameter to provide a sudden large disturbance to the system. This disturbance can cause disastrous tipping events in reality, as such abrupt jumps cannot be easily predicted~\cite{bifurcationdiagramsincomplexclimate,polarizationtippingpoints}. This shock or abrupt increase of control parameter is implemented in our laminar thermoacoustic system, Rijke tube. We conduct experiments where the voltage supplied to the grid is initially increased in a linear manner and then abruptly increased within a timespan of 0.3 ms to a value lying between $V_h$ and $V_f$ (within the bistable region of the system). The sudden increase in the voltage supplied to the grid is the shock applied to the control parameter, with a magnitude on the order of 1 V/ms. By positioning the final value of the voltage between $V_h$ and $V_f$, the system is brought into a regime where two attractors coexist, allowing us to examine how the sudden disturbance can cause the system to fall into the basin of attraction of LCO i.e., into a high-amplitude oscillatory state. The results of the experiments conducted when a shock is given to the control parameter are discussed in the following section.
\section{Effect of shock in control parameter on a thermoacoustic system} \label{exp_results}

\begin{figure}[H]
    \centering
    
    \begin{subfigure}[b]{0.49\textwidth}
        \centering
        \includegraphics[width=\textwidth]{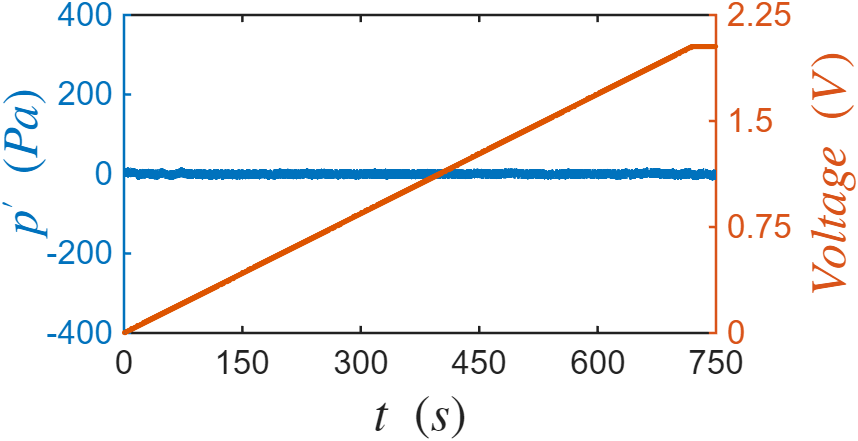}
        \caption{}
        \label{exp_quasi}
    \end{subfigure}
    \hfill
    \begin{subfigure}[b]{0.49\textwidth}
        \centering
        \includegraphics[width=\textwidth]{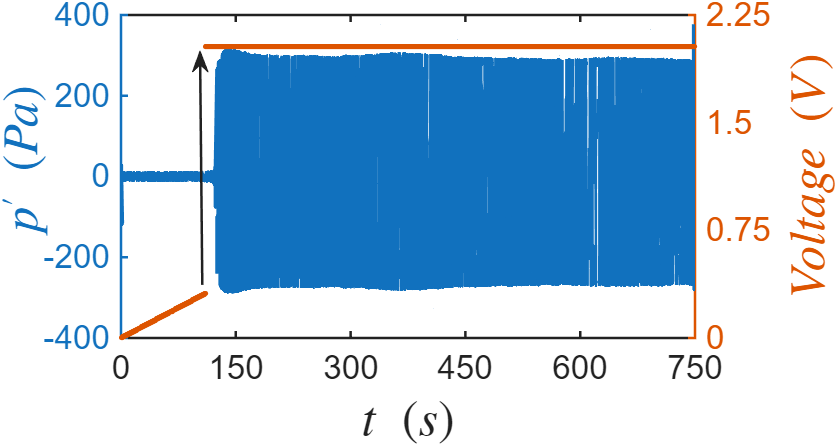}
        \caption{}
        \label{tai_exp}
    \end{subfigure}

    \caption{\large  (a) When the voltage is increased at a constant rate of 3~\text{mV/s}, within 720 s to $V_t = 2.16~\text{V}$, without a shock, the system remains in the quiescent state, reflected as the noise floor with a mean near zero. (b) The voltage is linearly increasing at a constant rate of 3~\text{mV/s} and at $\tau_t = 50$ s, the value of voltage is abruptly increased from $0.5~\text{V}$ to $2.16~\text{V}$ ($V_t$) as shown by the arrow, within $0.3$ ms and then maintained at this value until $750~\text{s}$. The shock induced drives the system from a quiescent state to high-amplitude periodic oscillations.}
    \label{exp_plots}
\end{figure}

\large  We initially perform quasi-static experiments to determine the Hopf and fold points of the system. When the control parameter is varied quasi-statically, the system transitions into a state of LCO from quiscent operation through a subcrtitical Hopf bifurcation~\cite{etikyala2017change}. Ten sets of quasi-static experiments were conducted from which the variation in the values of voltage corresponding to $V_h$ and $V_f$ are obtained to be 2.26 $\pm$ 0.04 V and 2.08 $\pm$ 0.04 V, respectively. 
\\
 \large We conduct two cases of experiments to understand the effect of shock in control parameter in the transition of the system to LCO. In the first case, experiments were conducted where the voltage supplied to the grid is linearly increased at a rate, $r$ = 3 mV/s, to a target value of $V_t$ = 2.16 V, situated between $V_h$ and $V_f$ and lies within the bistable region. Figure~\ref{exp_quasi} reveals that the system remains in its quiescent state when the control parameter, the voltage supplied to the grid is increased linearly and remained within the bistable region. Repeatability was ensured by performing 10 trials of the same experiment under identical conditions and confirming that the results remained the same.
\\
\large In the second case, we introduce an abrupt increase in the voltage supplied to the grid to take it to the target value ($V_t$) and the system tips from the quiescent state into a state of LCO which is illustrated in Fig.~\ref{tai_exp}. As mentioned in Section~\ref{exp_setup}, the voltage supplied to the grid increases linearly at $r = 3$ mV/s, and at a time instant $\tau_t = 50$ s a shock is applied by abruptly increasing the voltage from 0.50 V to $V_t = 2.16$ V ($\Delta V = 1.66$ V) within a short interval of $\Delta t = 0.3$ ms. The voltage is then held constant at $V_t$ for the remaining duration of the experiment. The system transitions from quiescent operation to a state of LCO,which is reflected as an increase in the acoustic pressure fluctuations from 4 Pa to 300 Pa. We perform 10 trials to verify repeatability and observe identical results across all trials.
\large

From the results of case 2 seen in Fig.~\ref{tai_exp} the mechanism of the tipping cannot be attributed to B-tipping, since the control parameter does not cross the Hopf point of the system. To investigate whether this transition could be driven by N-tipping, the case 1 shown in Fig.~\ref{exp_quasi} can be considered. Here, the control parameter is gradually increased and then held constant at the target voltage value ($V_t$, same as in the case in Fig.~\ref{tai_exp}). The corresponding acoustic pressure fluctuations of the system remains in a quiescent state, thereby ruling out the possibility that the noise in the system N-tipping could have triggered the tipping to LCO in case 2.
\\
\large Additionally, multiple experiments were performed by varying the initial rates of increase of the voltage supplied to the grid within the range \textit{r} = 3 -- 5 mV/s. Across all cases, we obtained similar results, indicating that the initial rate of increase does not influence the response of the system to the shock. Furthermore, the timescale over which the shock is applied (0.3 ms) is much shorter than the timescale associated with the initial rate of increase of the control parameter (1 s). We explore the dependence of the magnitude of shock (i.e., instantaneous change in the rate with time, $dr/dt$) using the mathematical model in Section~\ref{modelresults}.
\\
\large These observations collectively indicate that the onset of LCO is not governed by stochastic fluctuations (eliminating N-tipping), nor by the system crossing a bifurcation point (B-tipping). Therefore, the observed transition likely results from the mechanism of S-tipping, triggered by the abrupt increase in the voltage supplied to the grid.

\section{Mathematical model for the Rijke Tube} \label{math_model}
\large To decipher the possible mechanism behind the transitions seen in the experiments, we use a theoretical model capable of capturing the underlying phenomena. In Rijke tube~\cite{balasubramanian2008thermoacoustic,subramanian2010bifurcation}, applying the principles of momentum, Eq.~\eqref{momconsv} and energy conservation Eq.~\eqref{energyconsv} yields relationships between the acoustic velocity and the acoustic pressure. We remark that in this model, the acoustic wave propagation is modeled as linear, whereas the heat release rate is nonlinear. Neglecting the influence of mean flow and variations in the mean temperature, the governing equations for the acoustic field inside the duct are given by~\cite{balasubramanian2008thermoacoustic,subramanian2010bifurcation},
\begin{equation}
    \bar{\rho} \frac{\partial \tilde{u}'}{\partial \tilde{t}} + \frac{\partial \tilde{p}'}{\partial \tilde{x}} = 0 \quad 
\label{momconsv}
\end{equation}
\begin{equation}
    \frac{\partial \tilde{p}'}{\partial \tilde{t}} + \gamma \bar{\rho} \frac{\partial \tilde{u}'}{\partial \tilde{x}} = (\gamma - 1) \dot{\tilde{Q}}' \quad 
\label{energyconsv}
\end{equation}
\large All the variables with $\tilde{}$ (tilde)  represent dimensional variables and those without $\tilde{}$ (tilde) represent non-dimensionalized variables. The constants with dimensions are also expressed without $\tilde{}$ (tilde). $\bar{\rho}$ refers to the mean density of the air flow, $\tilde{u}'$ refers to the acoustic velocity perturbations, $\tilde{p}'$ refers to the acoustic pressure perturbations, $\gamma$ refers to the ratio of specific heats of air, $\dot{\tilde{Q}}'$ refers to the oscillatory heat release rate per unit area from the grid to the air flow around the grid. 
\\
\large The power supplied to the grid is dissipated~\cite{matveev2003thesis} as heat through convection, conduction, and radiation. Convection in the system can occur through natural, forced, or a combination of both mechanisms~\cite{beke2010modelling}. The rate of heating of the grid, which increases its temperature is governed by the net heat balance, determined by the electrical power supplied and the heat lost to the surroundings, as described by the governing heat transfer equation~\cite{matveev2003thesis}.
\begin{equation}
    \dot{\tilde{{Q}_g}} = \dot{\tilde{{Q}}}^{{fcon}}_{{ga}} + \dot{\tilde{{Q}}}^{{ncon}}_{{ga}} + \dot{\tilde{{Q}}}^{{cond}}_{{gr}} + \dot{\tilde{{Q}}}^{{rad}}_{{g}} - \tilde{{P}_g}
\label{matveeveq}
\end{equation}
\large Here subscript $g$ stands for the grid, $a$ stands for the air and $r$ for the rods connecting to the grid that supplies the electric power. $\dot{\tilde{{Q}_g}}$ refers to the rate of heating of the grid which increases its temperature, $\tilde{{P}_g}$ refers to the electric power supplied to the grid, ${\dot{\tilde{{Q}}}^{{fcon}}_{{ga}}}$ and \({\dot{\tilde{{Q}}}^{{ncon}}_{{ga}}}\)  refers to the heat transferred from the grid to the air through forced convection and natural convection respectively, \(\dot{\tilde{{Q}}}^{{cond}}_{{gr}}\) represents the heat transferred through conduction from the grid to the metal connecting rods which supplies the electric power and $\dot{\tilde{{Q}}}^{{rad}}_{{g}}$ refers to the heat transferred from the grid to the surroundings through radiation.
\\
\large Since convection can be natural or forced, their relative contributions are assessed using the Richardson number (\(Ri\)), which compares the buoyancy forces to inertial forces~\cite{beke2010modelling}. Natural convection is driven by buoyancy forces, while forced convection is governed by inertia forces.
\begin{equation}
    Ri = \frac{g \beta (\tilde{T_g} - T_0)d}{u_0^2}
\label{richardson}
\end{equation}
\large Here, $g$ is the acceleration due to gravity, equal to $9.8 \, \text{m/s}^2$; $\beta$ is the thermal expansion coefficient of the steel grid, valued at $10.8 \times 10^{-6} \, \text{W/mK}$; $\tilde{T_g}$ represents the temperature of the heated object, which in our study is the grid, taken to be at $693 \, \text{K}$; and $T_0$ denotes the ambient temperature, taken as $293 \, \text{K}$. The diameter $d$ of the wire grid is $1\times 10^{-3} \, \text{m}$, and $u_0$ is the mean bulk flow velocity, $(19.6 \pm 0.2) \times 10^{-2}\, \text{m/s}$. The magnitude of \(Ri\) determines the dominant mechanism~\cite{beke2010modelling}. When \( Ri < 0.1 \), natural convection effects are negligible, whereas for \( Ri > 10 \), forced convection effects are negligible. In the intermediate range \( 0.1 < Ri < 10 \), both mechanisms are significant. For the conditions in the Rijke tube, $Ri = 0.011 \pm 0.001$, indicating that natural convection is negligible. 
\\
\large The radiative heat transfer term is given by Eq.~\eqref{radeqn} 
\begin{equation}
\dot{\tilde{Q}}^{{rad}}_g = \sigma \, e \, A_g \left(\tilde{T_{g}}^4 - T_a^4\right),
\label{radeqn}
\end{equation}
\large where \(\sigma = 5.6 \times 10^{-8} \, \text{W/m}^2\text{K}^4\) is the Stefan–-Boltzmann constant, \(e = 0.3\) is the emissivity~\cite{Transmetra_Emissivity} and \(A_g = 2.4 \times 10^{-3} \, \text{m}^2\) is the effective surface area of the grid. Using the characteristic values of the temperature of the grid ($\tilde{T_g}$) and the air ($T_a$) as, \(T_{g} = 693\,\ \text{K}\) and \(T_a = 293\,\ \text{K}\), the calculated radiative heat transfer amounts to approximately 9 W.
This value is negligible compared to the other modes of heat transfer in the system, conduction ($\dot{\tilde{{Q}}}^{{cond}}_{{gr}}$) and forced convection(${\dot{\tilde{{Q}}}^{{fcon}}_{{ga}}}$), which are estimated to be approximately \(2.8 \times 10^2\, \text{W}\) and \(3.0 \times 10^4\, \text{W}\), respectively, based on the characteristic values substituted into Eq.~\eqref{conduction} and ~\eqref{fconv}.
 Given its comparatively minor contribution, the radiation term is considered insignificant and is therefore omitted from further analysis.
 Therefore, the supplied electrical power is primarily dissipated through forced convection to the airflow and conduction to the connecting rods given by Eq.~\eqref{fconv} and ~\eqref{conduction}.
\begin{equation}
    \dot{\tilde{{Q}}}^{{cond}}_{{gr}} = \frac{k A_r}{\Delta y} \left( \tilde{T_g} - {T_a} \right) 
\label{conduction}
\end{equation}
\large Here $k$ is the thermal conductivity, $\tilde{T_g}$ represent the temperature of the grid, the temperature of the connecting rods are taken to be equal to that of the air surrounding it (${T_a}$),  $A_r$ is the small area of contact  between the connecting rod and the grid and $\Delta y$ is the infinitesimal length of contact between the connecting rods and the grid.
\\
\large The heat release rate through forced convection from the heated grid to the airflow is reported~\cite{reynoldsnumberdependance} by Torszynski and O’Hern, \(\dot{\tilde{Q}}^{fcon}_{ga}\), given by Eq.~\eqref{fconv} :
\begin{equation}
    \dot{\tilde{{Q}}}^{{fcon}} = \left( \frac{\tilde{T_g} - T_a}{T_a} \right) \left( C \frac{\sqrt{\tilde{u}'}}{A_g} \right)\delta(\tilde{x} - \tilde{x_f})
\label{fconv}
\end{equation}
Where $C$ is a constant given by:
\begin{equation}
     C = \left( \frac{a_0^2}{\gamma - 1} \right) \left( \frac{(1 - k) A_g \mu}{1.01 \times D_g} \right) \left( \frac{1}{Pr^{0.6}} \sqrt{\frac{\rho D_g}{\mu}} \right)
\label{Cdef}
\end{equation}
\large In the above expression, $a_0$ is the speed of sound in air at room temperature, $\mu$ is the dynamic viscosity of air, ${k}$ is the ratio of the free area of the grid to the solid area, $D_g$ is the diameter of each wire component of the grid, $Pr$ is the Prandtl number, $\rho$ is the density of air, $\tilde{x}$ {and} $\tilde{x_f}$ are respectively the distance along the tube and the location of the grid within the duct from the decoupler end.
\\
\large The rate of heating of the grid can be expressed in terms of the mass of the grid ($m$), heat capacity ($C_p$) and the rate of change of its temperature, as follows:
\begin{equation}
    \dot{\tilde{{Q}_g}} = {m} C_p \frac{d \tilde{T_g}}{\tilde{d t}}
\label{gridheat}
\end{equation}
The governing heat transfer equation of the grid can be expressed as:
\begin{equation}
     \dot{\tilde{{Q}_g}} = \dot{\tilde{{Q}}}^{{fcon}}_{{ga}} + \dot{\tilde{{Q}}}^{{cond}}_{{gr}} - \tilde{{P}_g}
\label{heattransfer}
\end{equation}
The expression in Eq.~\eqref{heattransfer} can be rewritten using expansions of each term as follows:
\begin{equation}
    {m} C_p \frac{d \tilde{T_g}}{d\tilde{t}} = \left( \frac{\tilde{T_g} - T_a}{T_a} \right) \left( C \frac{\sqrt{\tilde{u}'}}{A_g} \right)\delta(\tilde{x} - \tilde{x_f}) + \frac{k A_r}{\Delta y} \left( \tilde{T_g} - T_a \right) - \tilde{P_g}
\label{heattransdiff}
\end{equation}
All the equations here are non-dimensionalized using the following terms:
\begin{equation}
\begin{aligned}
    x &= \frac{\tilde{x}}{L_{ac}}, \quad  
    \tau = \frac{\tilde{t}}{L_{ac}/a_0}, \quad
    u' = \frac{\tilde{u}'}{u_0}, \quad 
    p' = \frac{\tilde{p}'}{\bar{p}}, \\
    M &= \frac{u_0}{a_0}, \quad 
    T_g = \frac{\tilde{T_g}}{T_a}, \quad
    K = \frac{\tilde{P_g}}{C \cdot \sqrt{\tilde{u}} / A_g}
\end{aligned}
\label{nondim}
\end{equation}
The Mach number is denoted by \( M \), and the non-dimensional time by \( \tau \). The length of the tube which is the acoustic length scale is represented as \( L_{ac} \). The power supplied to the grid $P$ is non-dimensionalized using $C \cdot \sqrt{\tilde{u}}/A_g $ to get the equivalent non-dimensional  power supplied to the grid ($K$), where $C$ is as defined in Eq.~\eqref{Cdef}. The mean pressure within the tube is denoted by \( \bar{p} \). The velocity field, represented by \( \tilde{u} \), consists of both the mean flow velocity \( {u_0} \) and the acoustic perturbation component \( \tilde{u}' \), such that \( \tilde{u} = {u_0} + \tilde{u}' \).
Non-dimensionalization of Eq.~\eqref{heattransdiff} using the terms from Eq.~\eqref{nondim} yields:
\begin{equation}
    m C_p T_0 \frac{a_0}{L_a} \frac{d T_g}{d \tau} = \, C \left( T_g - 1 \right) \sqrt{u_0({1 + u^{\prime}})}
+ \, \frac{k A_r T_0}{\Delta y} \left( T_g - 1 \right) - K
\label{nondimheattransf}
\end{equation}
The rate of heat release from the grid to the air flow around it is equal to the rate of heat released through forced convection ($\dot{\tilde{{Q}}}^{{fcon}}$) and is given by Eq.~\eqref{fconv}. This can be substituted for $\dot{\tilde{Q}}'$  in the governing equation for acoustic energy conservation given by Eq.~\eqref{energyconsv} and non-dimensionalized to obtain:  
\begin{equation}
    \frac{\partial p'}{\partial \tau} + \frac{\gamma u_0}{a_0} \frac{\partial u'}{\partial x} = C_0 \left( {{T_g} - 1} \right) \sqrt{(1 + u')} \delta (x - x_f)
\label{energynew}
\end{equation}
where $C_0$ is a constant defined as :
\begin{equation}
    C_0 = \frac{(\gamma - 1)}{\bar{p} a_0} L_a \frac{C}{A_g} \sqrt{u_0}
\label{c0def}
\end{equation}
Equation ~\eqref{momconsv} is also non-dimensionalized using the quantities from Eq.~\eqref{nondim} to obtain the following equation:
\begin{equation}
    \gamma M \frac{\partial u'}{\partial \tau} + \frac{\partial p'}{\partial x} = 0
\label{momconsvnondim}
\end{equation}
Galerkin technique~\cite{balasubramanian2008thermoacoustic,zinnandlores} is used to convert Eqs.~\eqref{energynew} and~\eqref{momconsvnondim} to a set of ordinary differential equations.
This technique approximates the solution of differential equations using basis functions that satisfy the boundary conditions. These functions are not unique, and in this case, the self-adjoint part of the eigenfunctions of the linearized thermoacoustic system are selected. The 
acoustic velocity and pressure field expressed using the basis functions as follows:
\begin{equation}
    u^{\prime} = \sum_{j=1}^{\infty} \eta_j \cos(j \pi x) \quad \text{and} \quad p^{\prime} = - \sum_{j=1}^{\infty} \frac{\gamma M}{j \pi} \dot{\eta_j} \sin(j \pi x).
\label{modes}
\end{equation}
Here, $ \eta_j$ and $\dot{\eta_j}$ are the Galerkin modes of the system~\cite{zinnandlores}.
We substitute the first mode ($j = 1$) of the expansions in Eq.~\eqref{modes} into Eqs.~\eqref{energynew} and ~\eqref{momconsvnondim} and project them along the basis functions to reduce them to ordinary differential equations~\cite{balasubramanian2008thermoacoustic} (ODE);
\begin{equation}
    \frac{d \eta_j}{d \tau} = \dot{\eta_j}
\label{modmom}
\end{equation}
\begin{equation}
    \frac{d\dot{\eta_j}}{d \tau} + 2\eta_j \omega_j \zeta_j + k_j^2 \eta_j = -\frac{2jC_0 \pi}{{M}} \left( {{T_g} - 1} \right) \sqrt{1 + u'(t - \tau_0)} \sin(\pi j x_f)
\label{energymod}
\end{equation}
Where, $\tau_0$ is the time lag 
and the term $2 \eta_j \omega_j \zeta_j$ accounts for the frequency dependent damping in the system~\cite{subramanian2010bifurcation,KIMatveevmodel,Sterlingdamping}. Here $\zeta_j = {1}/{2\pi} \left[ c_1 {\omega_j}/{\omega_1} + c_2 \sqrt{{\omega_1}/{\omega_j}} \right]$, with $c_1$ and $c_2$ being constant coefficients taken to be equal to 1. For the first mode, $j$ = 1 for $\omega_j = \omega_1$, gives $\zeta = 0.16$.\\
The three ODE's Eq.~\eqref{heattransdiff}, ~\eqref{modmom} and ~\eqref{energymod} are solved using the 4$^{th}$ order Runge-Kutta method to obtain the values of acoustic pressure, velocity, and grid temperature.

\section{Model results for S-tipping in Rijke tube} \label{modelresults}
\begin{figure}[H]
    \centering
    
    \begin{subfigure}[b]{0.5\textwidth}
        \centering
        \includegraphics[width=\textwidth]{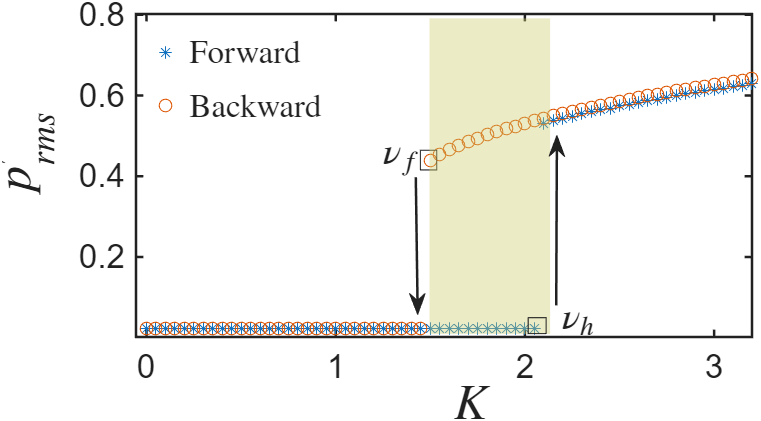}
        \caption{}
        \label{bif_model}
    \end{subfigure}
    \hfill
    \begin{subfigure}[b]{0.49\textwidth}
        \centering
        \includegraphics[width=\textwidth]{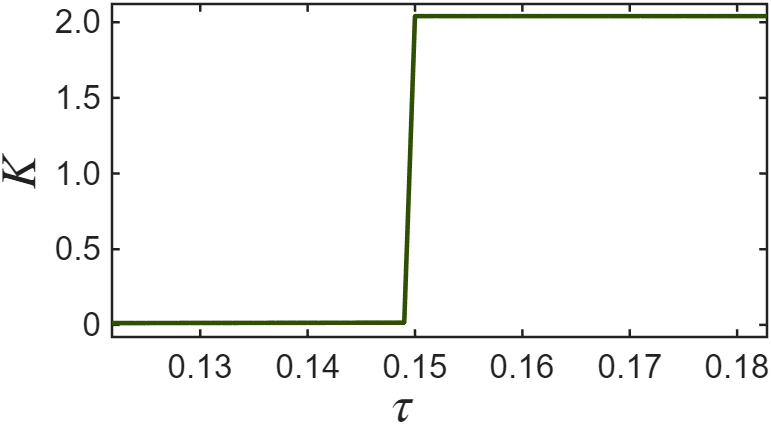}
        \caption{}
        \label{shock_model}
    \end{subfigure}

    \caption{ (a) \large Root mean square (RMS) values of the nondimensional acoustic pressure fluctuation $(p')$, obtained from the Rijke tube model, plotted against $K$.
The system jumps from quiescent state to the state of LCO, with a hysteresis region marked in yellow. The values of $K$ corresponding to the Hopf and fold points marked in the diagram as $\nu_h$ and $\nu_f$ are 2.05 and 1.45 respectively. (b) The variation of $K$ with $\tau$ which models the shock applied to the system. At a particular instant of $\tau$, denoted by $\tau_t$ = 0.15, the value of $K$ is increased within $10^{-3}$ units of nondimensional runtime from 0.015 to 2.04, giving a shock of magnitude, $dK/d\tau$ = 2025}.
\end{figure}

\begin{figure}[H]
    \centering
    
    \begin{subfigure}[b]{0.49\textwidth}
        \centering
        \includegraphics[width=\linewidth]{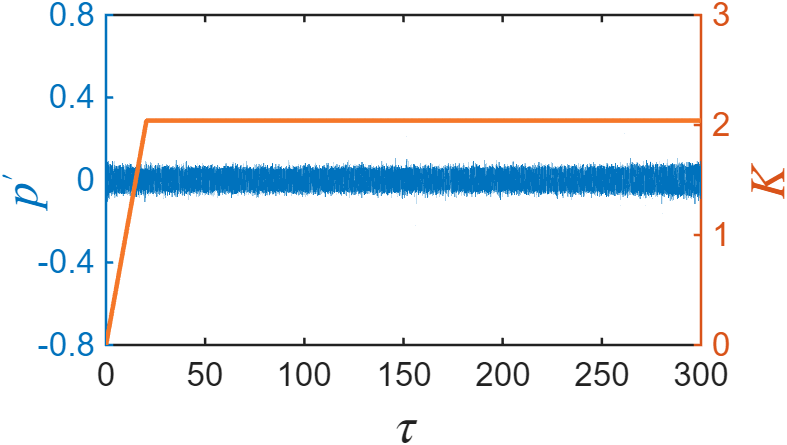}
        \caption{}
        \label{quasi_model}
    \end{subfigure}
    \hfill
    \begin{subfigure}[b]{0.49\textwidth}
        \centering
        \includegraphics[width=\linewidth]{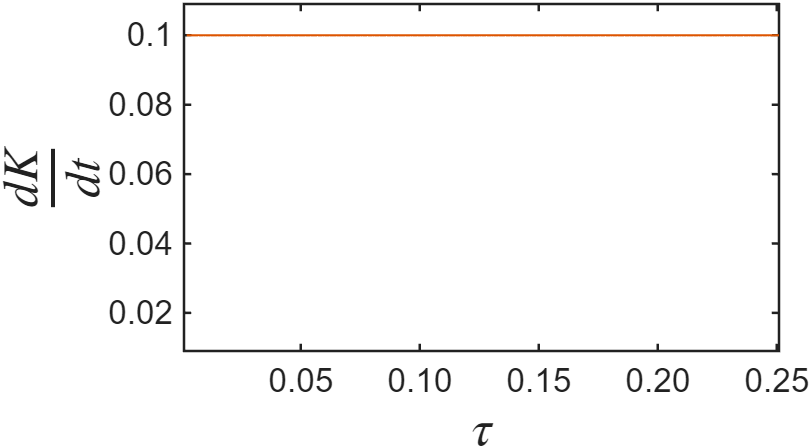}
        \caption{}
        \label{mag_shock_quasi}
    \end{subfigure}
    
    \vspace{0.5cm}
    
    \begin{subfigure}[b]{0.49\textwidth}
        \centering
        \includegraphics[width=\linewidth]{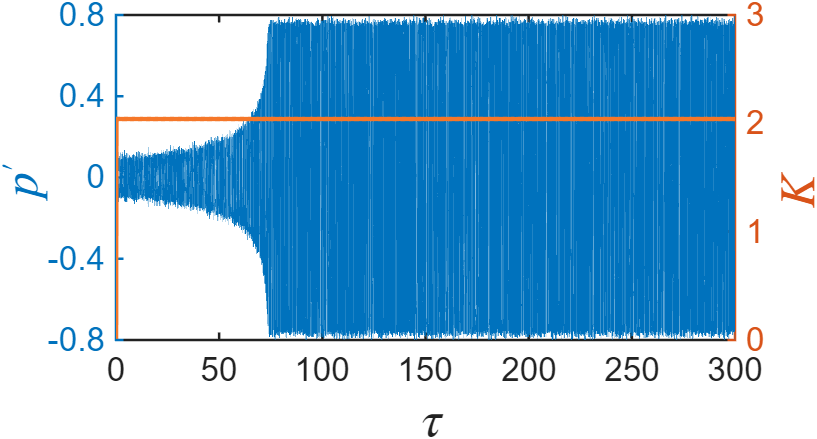}
        \caption{}
        \label{model_tai}
    \end{subfigure}
    \hfill
    \begin{subfigure}[b]{0.49\textwidth}
        \centering
        \includegraphics[width=\linewidth]{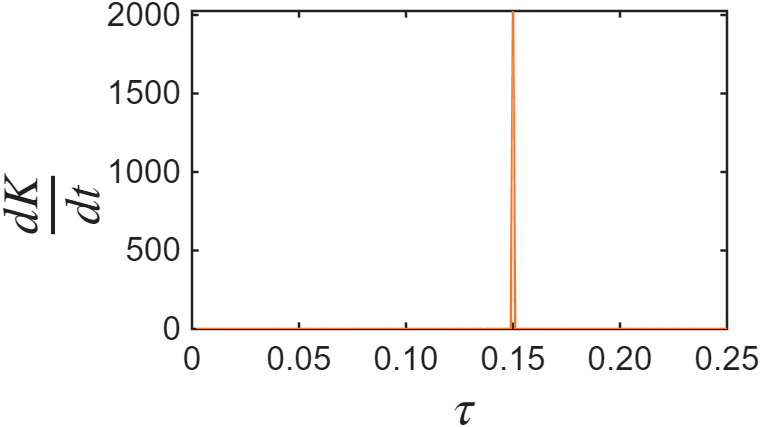}
        \caption{}
        \label{mag_shock_model}
    \end{subfigure}
    
    \vspace{0.5cm}
    
    \begin{subfigure}[b]{0.49\textwidth}
        \centering
        \includegraphics[width=\linewidth]{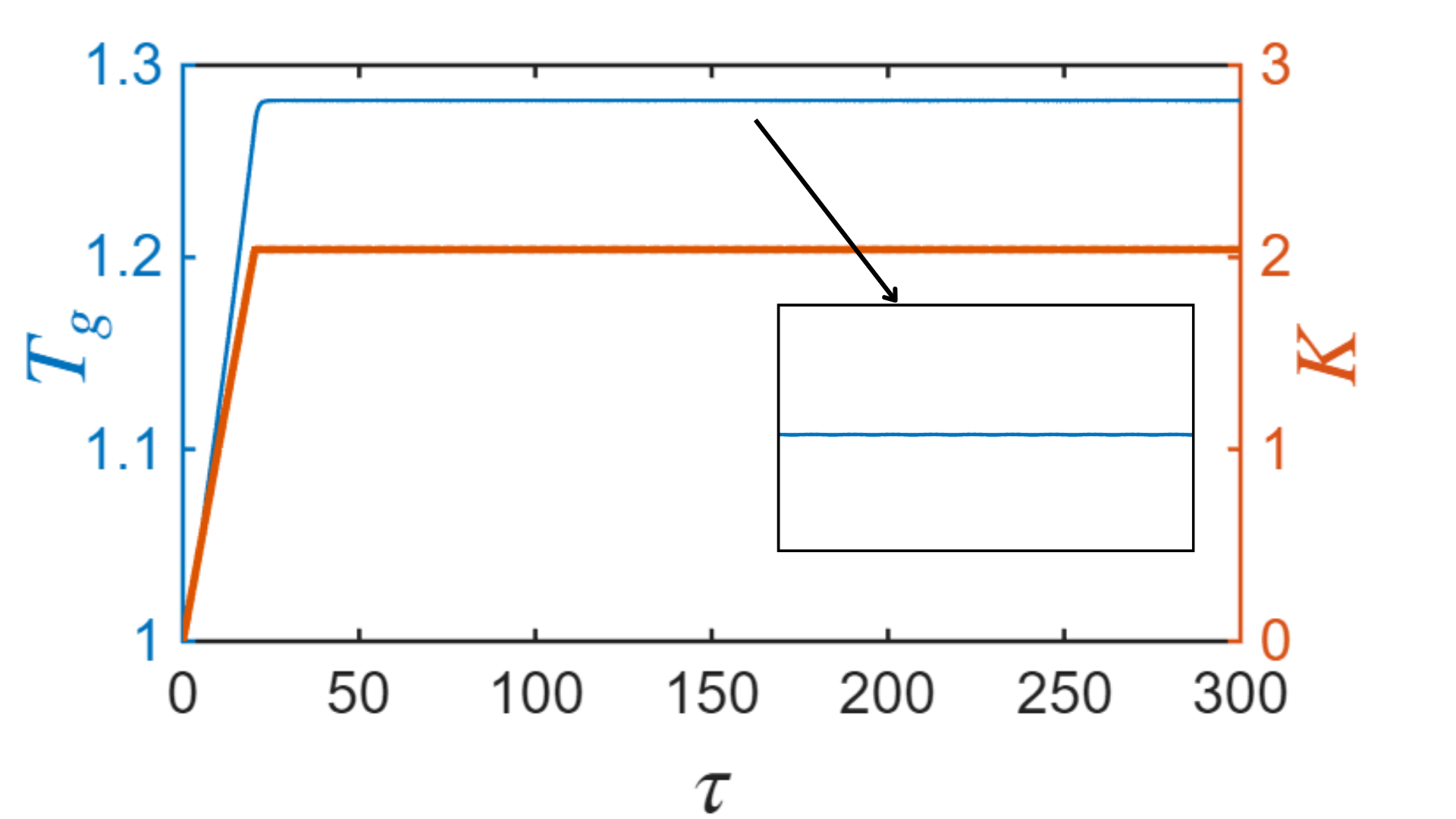}
        \caption{}
        \label{temp_time_quasi}
    \end{subfigure}
    \hfill
    \begin{subfigure}[b]{0.49\textwidth}
        \centering
        \includegraphics[width=\linewidth]{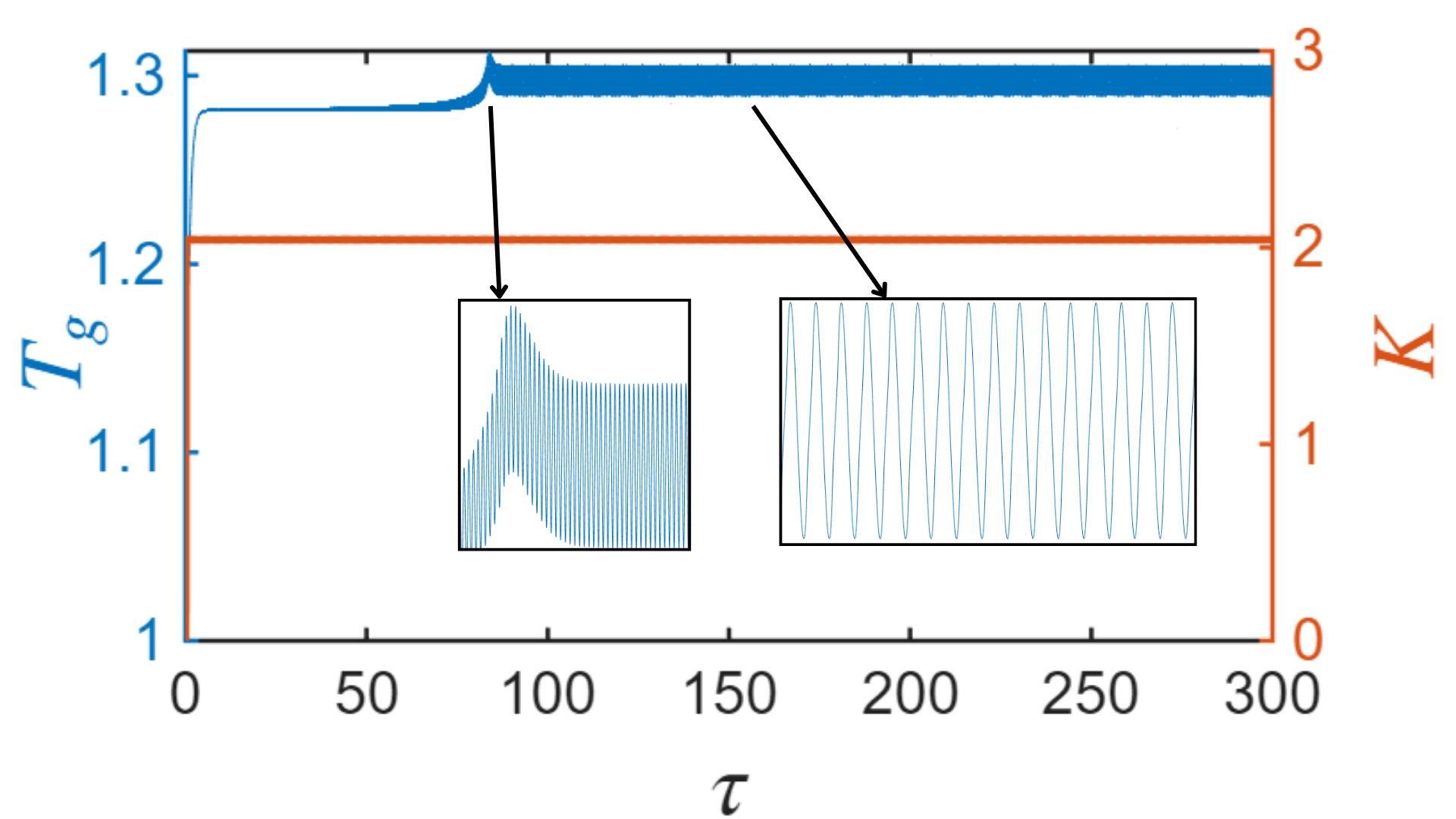}
        \caption{}
        \label{temp_time_tai}
    \end{subfigure}

\end{figure}

 The mathematical model presented in Section~\ref{math_model} is used to compute the solutions to the acoustic field and temperature variation of the grid in the Rijke tube system. For solving the coupled ODEs we use the initial conditions~\cite{balasubramanian2008thermoacoustic,subramanian2010bifurcation} \(\eta = 0.15\) and \(\dot{\eta} = 0.001\) and $T_0=297$K. The grid is placed one quarter of the length of the tube ($x_f = 0.25$), and the time lag \(\tau_0\) for heat release to the airflow~\cite{balasubramanian2008thermoacoustic,subramanian2010bifurcation,lighthill1954response} is set to 0.2. All results reported from the model are in non-dimensional form, with the control parameter defined as the non-dimensional equivalent power supplied to the grid ($K$). Increasing $K$, in steps of $10^{-3}$ to mimic quasi-steady experiments yields the Hopf ($\nu_h$) and fold ($\nu_f$) points of the system. As inferred from Fig.~\ref{bif_model}, these values are $\nu_h$ = 2.05 and $\nu_f$ = 1.45.
Figure ~\ref{shock_model} illustrates the variation of the control parameter when a shock is applied in the system. The value of \(K\), is initially linearly increased at a rate of \(r = 0.1\), and then abruptly raised at \(\tau_t = 0.15\) (as defined in Section~\ref{exp_results}) from \(K = 0.015\) to the target power point \(\nu_t = 2.04\). This value of \(\nu_t\), which lies between \(\nu_h\) and \(\nu_f\), i.e., the bistable region, is held constant for the remaining 300 units of non-dimensional runtime. 
\\
 The values of $\nu_t$ and $\tau_t$ are selected such that the magnitude of the shock, characterized by the derivative of the change in $K$ with respect to $\tau$, ${dK}/{d\tau}$, is of the order of $10^3$, which is found to be sufficient to drive the system into periodic LCO. We consider case 1, in which the control parameter is linearly ramped up to $\nu_t = 2.04$, a value between $\nu_f$ and $\nu_h$, is modeled first. As expected from the experiments, the system remains quiescent under this condition, as shown in Fig.~\ref{quasi_model}. For Case 2, where a shock is applied in the system to take it to $\nu_t$, the model shows that even though $\nu_t$ remains below $\nu_h$, the system still transitions into periodic LCO, as illustrated in Fig.~\ref{model_tai}.
\par \noindent
\begin{minipage}{\textwidth}
    \centering
    \addtocounter{figure}{-1}
       \captionof{figure}{\large(a) When \(K\) is linearly increased at a rate, $r$ = 0.1 to a value of \(\nu_t = 2.04\) the system continues to remain in the quiescent state. (b) The constant low value of ${dK}/{d\tau}$ =  0.1, in contrast to that in Fig.~\ref{mag_shock_model}, confirms that no shock is applied in this case. (c) The acoustic pressure fluctuations in the system resulting from the variation of the control parameter as shown in Fig.~\ref{shock_model}. (d) A sharp spike at \(\tau = 0.15\), indicates the application of a shock of magnitude of 2025. (e) When the control parameter is linearly increased to the target voltage value, the temperature of the grid ($T_g$) also follows a linear trajectory before approaching a final, constant value. (f) The variation of $T_g$ with time, in the case where shock is given to system. The applied shock in the control parameter \( K \), manifests as a sudden increase in the temperature of the grid at the point when the system tips into LCO.}
       \label{overall_figure_caption}
   \end{minipage}
   \par

\begin{figure}[H]
    \centering
    
    \begin{subfigure}[t]{0.49\linewidth}
        \centering
        \includegraphics[width=\linewidth]{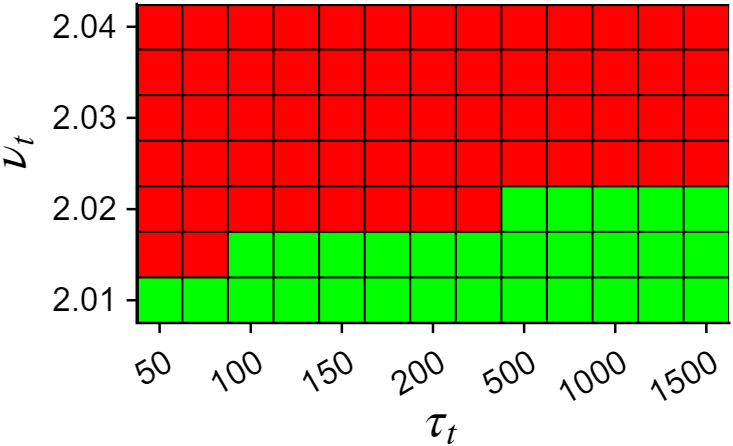}
        \caption{}
        \label{stabilitymap}
    \end{subfigure}
    \hfill
    \begin{subfigure}[t]{0.49\linewidth}
        \centering
        \includegraphics[width=\linewidth]{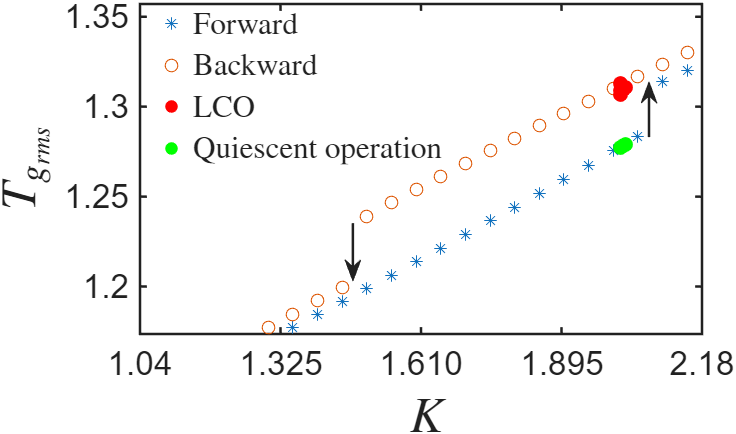}
        \caption{}
        \label{tbifspoints}
    \end{subfigure}

    \caption{(a) \large The stability map plotted for different $\nu_t$ and $\tau_t$ of the system. The red regions indicate the presence of LCO, while the green regions correspond to the quiescent state. We observe that as \(\tau_t\) increases, the system must be driven to a higher value of \(\nu_t\), i.e., closer to $\nu_h$ for the shock to tip the system to LCO. (b) The variation of $T_g$ with \(K\) is computed to construct the bifurcation plot with the grid temperature. The {red markers} denote the grid temperature at the exact point of tipping for the cases in Fig.~\ref{stabilitymap} where the system tipped to LCO. This transition occurred when the temperature exceeded the critical threshold $T_{\text{rms}} = 1.29$ for the combinations of $\nu_t$ and $\tau_t$ that lead to tipping to LCO in Fig.~\ref{stabilitymap}. In contrast, the {green markers} show the final grid temperature for cases that did not tip, as their temperatures remained below this critical threshold, preventing the onset of LCO.}
\end{figure}

\bigskip
The transition to LCO in Fig.~\ref{model_tai}, occurs without the system crossing its Hopf bifurcation point and can be attributed to the shock induced in $K$. 
A likely mechanism behind S-tipping lies in the heat transfer characteristics of the grid. As detailed in Section~\ref{math_model}, the grid temperature is an important variable in the system and plays a central role in governing the heat transfer within the Rijke tube. Direct temperature measurements in the systems cannot be captured accurately using thermocouples, as it cannot respond fast enough to the variations in the temperature of the grid. Hence the temperature variations are instead analyzed through the model to establish the hypothesis. The grid temperature obtained from the mathematical model presented in Section~\ref{math_model}, when plotted against $\tau$, shows a sharp rise precisely at the point where the pressure signal transitions into LCO, as illustrated in Fig.~\ref{temp_time_tai}.  In contrast, in the absence of a shock, this sudden temperature rise is not observed, as illustrated in Fig.~\ref{temp_time_quasi}. Even though the shock is applied to the voltage supplied to the system, an emergence of periodic behavior occurs in $T_g$. This indicates that even though the control parameter has not crossed the Hopf point, another variable of the system is getting triggered with such a shock in the control parameter. These findings highlight the critical role of abrupt temperature changes in initiating LCO in the system.
\\
Now that we have shown that a shock can drive the system from a quiescent state to LCO, it is essential to examine whether every time we give a shock, the outcome remains the same. The magnitude of a shock, as previously discussed, is mathematically expressed as \( dK/d\tau \), but in physical terms, it depends on the values of the target power value $\nu_t$ and the particular time instant when the shock is applied, $\tau_t$.  
\\
The relationship between the magnitude of the shock and the response of the system is influenced by the proximity of $\nu_t$ to the Hopf point. When $\nu_t$ is close to the Hopf point, only a low magnitude shock is required to tip the system into instability. However, when $\nu_t$ is farther from the Hopf point, a shock of much higher magnitude is needed to induce the transition. The influence of $\tau_t$ on the magnitude of the shock is inversely proportional; therefore, when $\tau_t$ is lower, the system requires a larger magnitude of shock, to tip into the state of LCO. This relationship can be effectively visualized using a stability map, as shown in Fig.~\ref{stabilitymap}, which illustrates the emergence and non-emergence of LCO in the system for different conditions of $\nu_t$ and $\tau_t$. 

 Figure ~\ref{tbifspoints} presents the variation of the root mean square (RMS) values of $T_g$ as the non-dimensional equivalent power supplied to the grid \( K \) is varied quasi-steadily. The bifurcation plot shows the role of temperature, an important variable of the system, in the onset of LCO. The $T_g$ values at the tipping points is plotted for different conditions of \( \tau_t \) and \( \nu_t \), for the cases shown in the stability map, Fig.~\ref{stabilitymap} is marked in bifurcation plot. In all cases where the system transitions into LCO, the grid temperature at tipping exceeds the critical value of \( T_{\text{rms}} = 1.29 \). This confirms that the onset of LCO is triggered by the grid crossing this critical temperature value which can be considered analogous to the system crossing the Hopf point in the bifurcation diagram for pressure. 
The shock drives the system from a fixed point to the bistable regime, where it falls into the basin of attraction of an alternate stable state. Although the control parameter remains below its value at the stability boundary, the system tips into the basin of attraction of the state of LCO when the system crosses the basin boundary in another variable which is here the temperature of the grid.
\\
We experimentally demonstrate that the Rijke tube system can tip into the state of LCO, when subjected to a shock in the voltage supplied to the grid. When the voltage supplied to the grid was gradually increased to a value within the bistable region, the system remained in the quiescent state. However, when the same final voltage was reached by applying a sudden shock producing a 1.66 V change within 0.3 ms, tipping to LCO was observed. These experiments reveal that shocks constitute a distinct mechanism for inducing tipping between states in a thermoacoustic system.
\\
Through a mathematical model, we successfully capture the mechanism behind the S-tipping in a Rijke tube. A shock in the control parameter produces a corresponding shock in the temperature of the grid, which pushes the system into the basin of attraction of the state of LCO. The bifurcation diagram of the grid temperature reveals the existence of a critical temperature threshold beyond which the system tips to the state of LCO. This study brings out the importance of identifying auxiliary system variables, beyond the primary control parameter, whose crossing of a critical threshold can irreversibly redirect the system into a different basin of attraction and drive the system into an alternate dynamical state.

\section{Conclusions}
\large In this study, shock-induced tipping (S-tipping) was investigated in a prototypical laminar thermoacoustic system, a horizontal Rijke tube. The experiments demonstrate that a sudden shock in the control parameter can drive the system from a quiescent state to high-amplitude LCO, even when the values of the control parameter values stay within the stability bounds. Theoretical modeling provided strong support to the experimental findings by showing that when the system is taken to the bistable region due to the shock, the system undergoes a transition from a quiescent state to LCO. The basin boundary is crossed in another variable of the system, which is the temperature of the grid, causing the system to fall into the basin of attraction of the state of LCO. This work establishes that shock-induced tipping can occur in a simple thermoacoustic system and highlights the role of sudden changes in control parameters in causing undesirable transitions.
\\
 \large Understanding S-tipping in such practical systems has broader implications for predicting and controlling tipping in various natural and engineering systems. In real-world engineering systems, such abrupt shifts in control parameters may occur due to sensor malfunctions~\cite{carsgonewild}, supply surges~\cite{powersurges}, or unexpected operational failures~\cite{flightfailure}. Such events can inadvertently lead to catastrophic transitions. By understanding the mechanism behind S-tipping in a practical setup, our model provides valuable insight for system designers and engineers, emphasizing the need for prevention strategies. Anticipating such tipping behavior by being aware of the stability boundaries of the variables in the system can help in avoiding undesirable operational states. This, in turn, helps ensure safer and more reliable system performance across a wide range of applications.

\section{Acknowledgments}
We acknowledge the funding from IoE-IITM Research Initative \\(SP/22-23/1222/CPETWOCTSHOC). R. R acknowledges the support from Prime Minister’s Research Fellowship, Govt.\ of India.
The authors gratefully acknowledge Dr.\ Induja Pavithran  for her valuable suggestions and critical comments.
The authors would like to thank Mr. Sridhara Tirumala Durga for his support with experiments.
Authors would also like to thank Mrs.\ Sudha G for the help provided in conducting the experiments. 

\section{Author Declarations}
\textbf{Conflict of Interest}
The authors have no conflicts to disclose.
\textbf{Data availability}
The data that support the findings of this study are available
from the corresponding author upon reasonable request.

\bibliographystyle{unsrtnat}

\bibliography{references}

\end{document}